\documentclass[aps,prc,superscriptaddress,showpacs,floatfix]{revtex4}

\usepackage[utf8]{inputenc}
\usepackage{amssymb}
\usepackage{amsmath}
\usepackage{hyperref}
\usepackage{graphicx}
\usepackage{multirow}
\usepackage{bm}
\usepackage{verbatim}
\begin{document}
\title{Implication of two-baryon azimuthal correlations in $pp$ collisions at LHC energies on the QGP}

\author{Liuyao Zhang}
\affiliation{Key Laboratory of Nuclear Physics and Ion-beam Application (MOE), Institute of Modern Physics, Fudan University, Shanghai 200433, China}
\author{Jinhui Chen}
\affiliation{Key Laboratory of Nuclear Physics and Ion-beam Application (MOE), Institute of Modern Physics, Fudan University, Shanghai 200433, China}
\author{Wei Li}
\affiliation{T.W. Bonner Nuclear Laboratory, Rice University, Houston, Texas 77005, USA}
\author {Zi-Wei Lin}
\affiliation{Department of Physics, East Carolina University, Greenville, North Carolina 27858, USA}
\date{\today}

\begin{abstract}
The near-side depression in two-proton or two-antiproton azimuthal correlations in $pp$ collisions at $\mathrm{\sqrt{s}= 7}$ TeV has been observed experimentally and then qualitatively reproduced in our earlier studies with a multi-phase transport model. In this study, we further investigate the origin of the depression feature in two-baryon correlations in small collision systems. We find that the initial parton-level spatial correlation,
a finite expansion in the parton stage, and quark coalescence are important ingredients leading to the near-side depression. In particular, we find that a finite expansion of the parton system leads to a finite space-momentum correlation at hadronization, which then converts the near-side depression in the coordinate space to that in the momentum space. These results suggest that a partonic matter with a finite lifetime is formed in the $pp$ collisions. 
Further studies are needed to determine whether the partonic matter is near local equilibrium and can thus be called a QGP or far away from local equilibrium.
\end{abstract}
\maketitle

\section{Introduction}

Particle correlations are powerful tools for the study of particle production and dynamics of relativistic nuclear collisions. Two-particle correlations are often presented in the $\Delta\eta$-$\Delta\varphi$ space with $\Delta\eta$ and $\Delta\varphi$ referring to the pseudorapidity difference and azimuth difference between the two hadrons in each pair, respectively. The most significant feature of multi-particle correlations in heavy ion collisions is due to the elliptic flow, an azimuthal anisotropy in momentum space~\cite{I.Arsene:2005,B.B.Back:2005,J.Adams:2005,K.Adcox:2005}
that is generated from the pressure gradient of the quark-gluon plasma (QGP) in the hydrodynamical evolution~\cite{Sirunyan:2017uyl} or particle scatterings in transport models~\cite{Lin:2001zk} in the initial almond-shaped overlap area of two nuclei. 
However, similar long-range azimuthal correlations have also been observed in small collision systems at high multiplicities~\cite{Abelev:2009af,Khachatryan:2010vs,Aad:2013,Chatrchyan:2013,Abelev:2013,Adare:2015,Aad:2015gqa}, where a consensus on its origin has not been reached yet~\cite{He:2015hfa,Lin:2015ucn,Dusling:2015gta,Zhao:2020}. Note that similar studies have been performed in $ee$ and $ep$ collisions~\cite{Badea:2019vey,ZEUS:2019jya}, where a ridge structure was not observed, and experimental upper limits were set on the possible magnitude of flow harmonic coefficients. 

An interesting feature in two-particle correlations is the near-side anticorrelation structure observed in baryon-baryon correlations along the $\Delta\varphi$ direction at low transverse momentum ($p_{T}$) in $pp$ collisions at $\rm \sqrt{s}$= 7 TeV~\cite{J.Adam:2017}. 
A similar depression had been reported in $e^{+}e^{-}$ collisions at 29 GeV and was attributed to fragmentation~\cite{Althooff:1984, Aihara:1985, Aihara:1986fy,Abreu:1993}. However, fragmentation alone can not explain the new experimental data from $pp$ collisions at $\mathrm{\sqrt{s}= 7}$ TeV~\cite{J.Adam:2017}. In our previous studies, the string melting version of a multi-phase transport (AMPT) model~\cite{Lin:2001zk} with an improved quark coalescence has been shown to qualitatively describe the anticorrelation feature in baryon-baryon correlations at LHC energies~\cite{Zhang:2018ucx,Zhang:2019bkf}. Here we further investigate the physics mechanism responsible for the depression feature including the initial parton-level spatial correlations, parton evolution and the quark coalescence. 

\section{Model}

A multi-phase transport model~\cite{Lin:2004en} is used in this study. In this model, the initial conditions are taken from the minijet partons and soft string excitations from the ${\rm HIJING}$ event generator~\cite{Wang:1991hta}. Then the two-body elastic parton scatterings are modeled with the parton cascade model ${\rm ZPC}$~\cite{Zhang:1997ej}. The conversion from partons to hadrons is performed via either the Lund string fragmentation for the default version or a quark coalescence model for the string melting version~\cite{Lin:2004en}. Hadronic interactions are treated with an extended version of a relativistic transport model ${\rm ART}$~\cite{Li:1995}. The string melting version of the AMPT model (AMPT-SM)~\cite{Lin:2001zk} is applicable when the parton degrees of freedom are expected in the initial state. For large systems at high energies, the model can simultaneously describe the pion and kaon rapidity distribution, $p_{T}$ spectra, and elliptic flow at low $p_{T}$~\cite{Lin:2014tya,Ma:2016fve}; it can also provide valuable information to understand the vector meson spin alignment signals~\cite{Shen:2021pds,Lan:2017nye}. 

In addition, the model is able to describe the anisotropic flows in small collision systems such as $p$Pb collisions at the LHC~\cite{Bzdak:2014dia,Wang:2022fwq}.
The recent improvement of the quark coalescence model for hadronization in the AMPT model~\cite{He:2017tla} improves the description of baryons and antibaryons~\cite{He:2017tla,Shao:2020}. 
In this new quark coalescence model, a parton has the freedom to form either a meson or a baryon depending on the relative distance to its coalescing partners; this is more physical because it removes the artificial constraint in the previous quark coalescence model that forces the numbers of mesons, baryons, and antibaryons in an event to be separately conserved through the quark coalescence process. Compared to earlier versions of the AMPT model, it also provides a better description of two-baryon correlations in $pp$ collisions at $\sqrt{s}$=7 TeV~\cite{Zhang:2018ucx,Zhang:2019bkf}; we thus use the same version for this study.

\section{Results}

Our analysis method for two-particle correlations is identical to that used in the $pp$ experiment~\cite{J.Adam:2017}. The two-particle correlation function $C(\Delta\eta,\Delta\varphi)$
is defined as the ratio of the same event distribution $S(\Delta\eta,\Delta\varphi)$ over the combinatorial distribution $B(\Delta\eta,\Delta\varphi)$. 
Here, $B(\Delta\eta,\Delta\varphi)$ is the distribution of particle pairs with one particle in one event and the other particle in another event with a similar multiplicity, while 
both the $S$ and $B$ correlation functions are properly normalized~\cite{J.Adam:2017, Zhang:2018ucx,Zhang:2019bkf}.
From the two-dimensional correlation function $C(\Delta\eta,\Delta\varphi)$,
an one-dimensional $\Delta\varphi$ correlation function $C(\Delta\varphi)$ is constructed after integrating both $S$ and $B$ correlation functions over a given $\Delta\eta$ interval~\cite{Zhang:2018ucx}. 

\begin{figure*}[htb]
\centering
\includegraphics[scale=0.7]{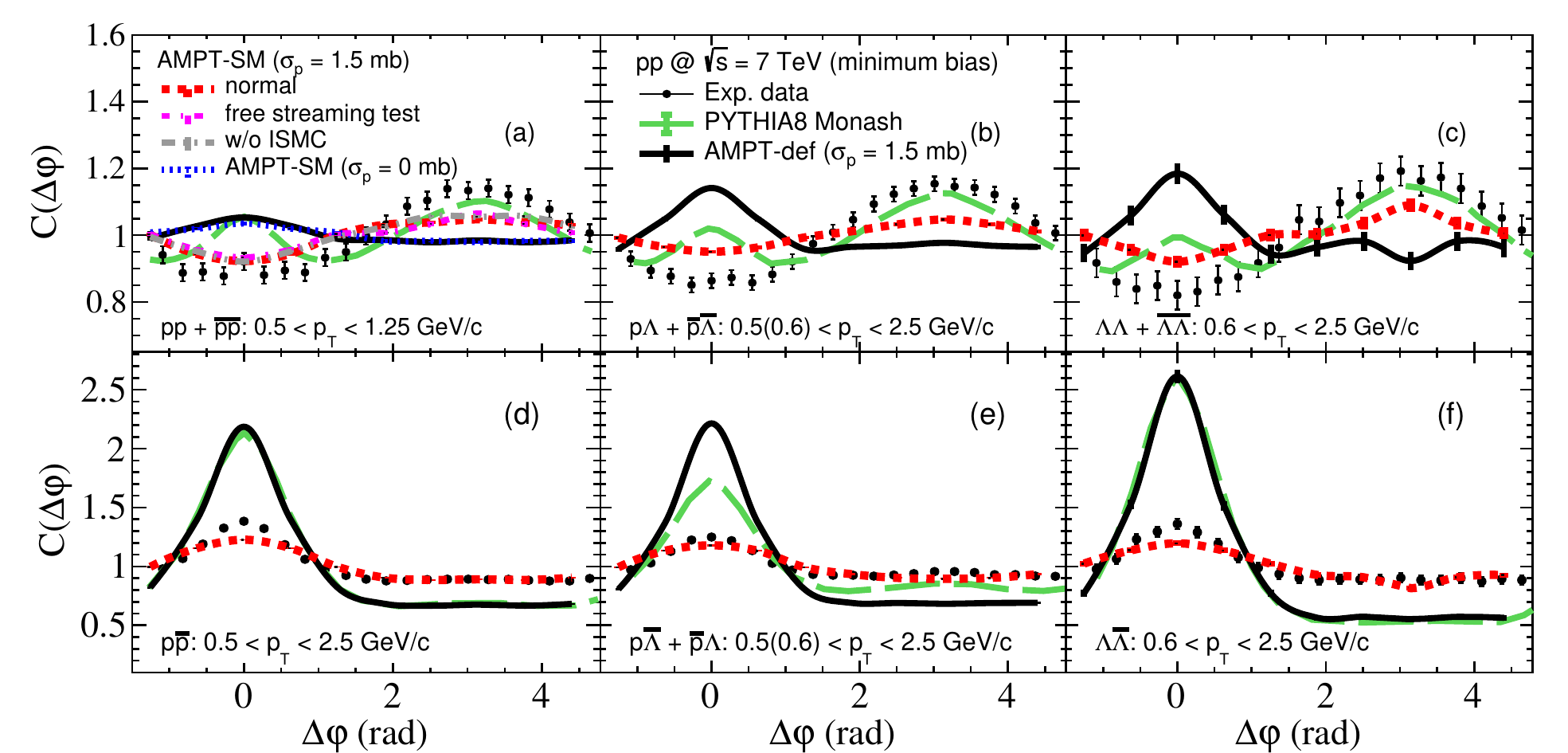}
\caption{(color online) Azimuthal correlations of (a) $pp+\bar{p}\bar{p}$, (b) $p\Lambda+\bar{p}\bar{\Lambda}$, (c) $\Lambda\Lambda+\bar{\Lambda}\bar{\Lambda}$, (d) $p\bar{p}$, (e) $p\bar{\Lambda}+\bar{p}\Lambda$, and (f) $\Lambda\bar{\Lambda}$ in minimum bias $pp$ collisions at $\rm \sqrt{s} = 7$ TeV from the default AMPT model and the string melting AMPT model at 1.5 mb parton cross section for the normal configuration, the free streaming test, and a test without the initial state momentum correlation. Filled circles represent the experimental data, long dashed curves represent the PYTHIA8 results~\cite{J.Adam:2017}, 
and the dotted curve in panel (a) represents the string melting AMPT model at 0 parton cross section.}
\label{fig1}
\end{figure*}

Figure~\ref{fig1}(a) shows the $\rm pp +\bar{p}\bar{p}$ (i.e., two-proton as well as two-antiproton) azimuthal correlation for minimum bias $pp$ collisions at $\sqrt{s}$ $=$ 7 TeV from the AMPT model with different configurations. 
We focus on $\rm pp +\bar{p}\bar{p}$ correlations in this study, where protons or antiprotons are selected within $|\eta|<0.8$ and 0.5 $\rm <p_T<$ 1.25 GeV/c for both the model results and data if not specified otherwise. 
We also show other like-sign baryon correlations including $p\Lambda+\bar{p}\bar{\Lambda}$ in Fig.~\ref{fig1}(b) and  $\Lambda\Lambda+\bar{\Lambda}\bar{\Lambda}$ in Fig.~\ref{fig1}(c), where protons or antiprotons ($\Lambda$s or $\bar{\Lambda}$s) are within 0.5 (0.6)$< p_{T} <$ 2.5 GeV/c for both the experimental data and model. 
We see that the string melting version of the AMPT model at parton cross section $\rm \sigma_p$=1.5 mb shows the anticorrelation features in the like-sign baryon correlations, consistent with our previous results~\cite{Zhang:2018ucx}. 
We also see that the anticorrelation is not present in either PYTHIA8~\cite{Sjostrand:2007gs,Skands:2014pea}
or the default version of AMPT (AMPT-def), which means that neither the fragmentation process alone 
nor fragmentation plus hadron cascade can explain the anticorrelation in the data. 
In addition, Fig.~\ref{fig1}(a) shows that the AMPT-SM result at 0 mb parton cross section does not show the anticorrelation feature either, suggesting that the parton phase is necessary for the depression feature.

We have also investigated the $\rm pp +\bar{p}\bar{p}$ correlation from the AMPT-SM model with two other configurations. 
First, we remove the initial state momentum correlation (ISMC) by randomizing the azimuthal angle of each parton's initial momentum while keeping all other variables (such as the initial $p_T$, $p_z$, position and time upon formation) unchanged. 
As shown in Fig.~\ref{fig1}(a), the result of the AMPT model without ISMC is very similar to the normal AMPT result, indicating that the initial state momentum correlation in the AMPT-SM model is unimportant for its depression feature in $pp$ collisions at $\rm \sqrt{s} =$ 7 TeV. 
We note that ISMC in the color glass condensate approach often plays an important role in final hadron momentum anisotropies in small systems~\cite{Giacalone:2020byk}. 
Second, we perform a free-streaming test, where the parton cross section is the same as the normal AMPT-SM case but the scattering angle is set to zero for every parton scattering. The $C(\Delta\varphi)$ correlation from the free-streaming test, as shown in Fig.~\ref{fig1}(a), is similar to the normal AMPT-SM result. 
For completeness, we also show the unlike-sign baryon correlations in Figs~\ref{fig1}(d-e), where results from the AMPT-SM model are much closer to the experimental data than results from the AMPT-def model or PYTHIA8. 

\begin{figure*}[htb]
\centering
\includegraphics[scale=0.7]{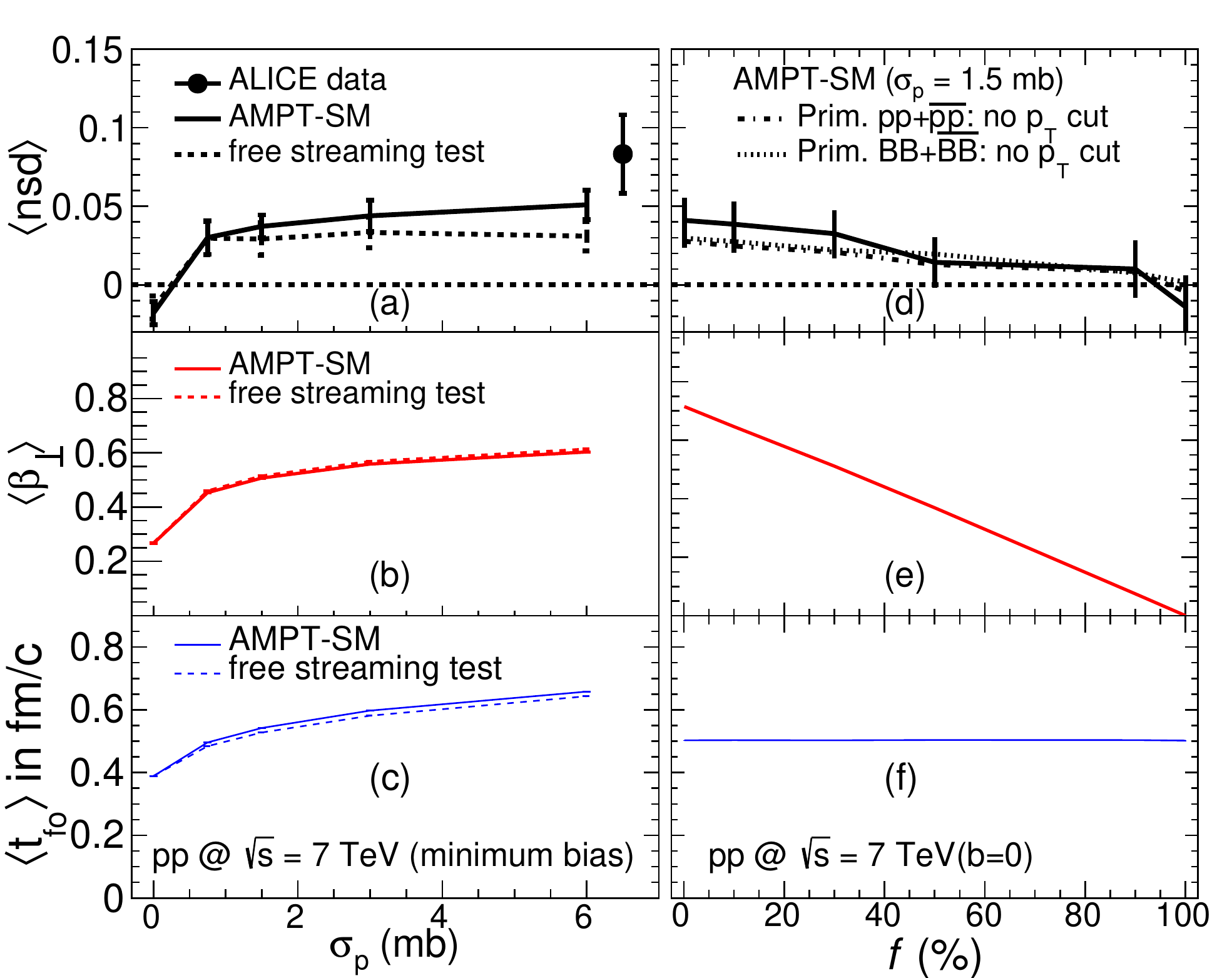}
\caption{(color online) The average near-side depression in baryon-baryon (and antibaryon-antibaryon) azimuthal correlations from the AMPT-SM model for (a) minimum bias $pp$ collisions at $\mathrm{\sqrt{s}= 7}$ TeV at different parton cross sections and (d) $b=0$ collisions after azimuthal randomization of different fractions of partons before coalescence. Both model results and experimental data use hadrons within 0.5 $\rm <p_T<$ 1.25 GeV/c. The corresponding average transverse space-momentum correlation $\langle\beta_{\perp}\rangle$ are shown in panels (b) and (e) and the corresponding freezeout times $\langle\mathrm{t_{fo}}\rangle$ are shown in panels (c) and (f) for freezeout partons at midrapidity.
}
\label{fig2}
\end{figure*}

To quantify the near-side depression (nsd), we define the average near-side depression of the correlation function $C(\Delta\varphi)$ from unity as
\begin{equation}
\langle \mathrm{nsd} \rangle \equiv 1-\frac{1}{\pi}\int^{+\pi/2}_{-\pi/2}C(\Delta\varphi) d\Delta\varphi.
\end{equation}
Figure~\ref{fig2}(a) shows the $\rm pp +\bar{p}\bar{p}$ $\langle \mathrm{nsd} \rangle$ results from AMPT model for different parton cross sections, while Fig.~\ref{fig2}(b1) shows the value extracted from data in minimum bias pp collisions at $\mathrm{\sqrt{s}= 7}$ TeV. We see that the near-side depression from the normal AMPT-SM model (black solid curve) first increases significantly with $\rm \sigma_p$, changing from a negative value 
(i.e., an enhancement instead of a depression) at $\rm \sigma_p=0$ mb to a positive
value at $\rm \sigma_p=0.75$ mb. It then tends to saturate at larger $\rm \sigma_p$, although the $\langle \mathrm{nsd} \rangle$ magnitude is lower than that observed in the experiment. 

To further investigate the cause of the depression feature, 
we calculate the space-momentum correlation variable in the transverse plane as~\cite{He:2015hfa}
\begin{equation}
\langle \beta_{\perp} \rangle 
\equiv \langle \hat {\mathbf{r}}_{\perp} \cdot \hat {\mathbf{p}}_{\perp} \rangle
=\left \langle \frac {{\mathbf{r}}_{\perp} \cdot {\mathbf{p}}_{\perp}} {r_{\perp} ~p_{\perp}} \right \rangle. 
\end{equation}
In the above, the bracket represents the averaging over particles in the events, 
${\mathbf{r}}_{\perp}$ and ${\mathbf{p}}_{\perp}$ represent the parton transverse position and momentum vector, respectively, where the origin in the transverse plane is the middle point  between the two incoming protons. The $\langle \beta_{\perp} \rangle$ values of mid-rapidity partons at freezeout from the normal AMPT-SM model as a function of the parton cross section are shown in Fig.~\ref{fig2}(b) (solid curve), 
which increase significantly with $\rm \sigma_p$ similar to $\langle \mathrm{nsd} \rangle$.
In addition, the magnitude of $\langle \beta_{\perp} \rangle$ is closely related
to the average freezeout time of mid-rapidity partons (blue solid curve). 
This is natural because a larger parton cross section $\rm \sigma_p$ delays the hadronization in the AMPT-SM model and a longer-lived parton phase leads to stronger space-momentum correlation of partons at freezeout.

These results bring up an interesting question: is the parton cross section or space-momentum correlation responsible for the observed depression feature? We emphasize that space-momentum correlation and parton interaction-induced collectivity (similar to the hydrodynamic collective flow for large systems at high energies)
are different~\cite{He:2015hfa}. For example, free streaming of partons from a compact source will soon generate a positive space-momentum correlation. 
Similarly, free streaming in hydrodynamics has been shown to lead to a radial flow similar to or even stronger than that from normal hydrodynamics~\cite{Romatschke:2015dha}. To check this, we design the free-streaming test   
where the parton cross section is the same as the normal AMPT-SM case but the scattering angle is set to zero for every parton scattering. As a result, no parton changes its three-momentum, but the finite $\rm \sigma_p$ enables the parton system to expand to a finite size and then kinetically freeze out like the normal case. 
As shown in Fig.~\ref{fig2}(a-c), results from the free-streaming test (dashed curves), including the average depression, space-momentum correlation and parton freezeout time, are quite similar to the normal AMPT-SM results. 
The $C(\Delta\varphi)$ correlation from the free-streaming test in Fig.~\ref{fig1} is also similar to the normal AMPT-SM result. The similarity between the normal case and the free-streaming test 
indicates that an essential ingredient responsible for the baryon pair depression is a finite expansion of the parton system till hadronization, which introduces a finite space-momentum correlation at hadronization. On the other hand, the strength of the parton interaction is not important for the depression feature.

To further demonstrate the connection between the space-momentum correlation and the near-side depression, we perform another test by randomizing 
the transverse momentum direction (i.e., the $\varphi$ angle) of a fraction ($f$) of the freezeout partons and then evaluating the $\langle \mathrm{nsd} \rangle$ of the primordial (anti)baryons after quark coalescence. 
Note that here we evaluate primordial hadrons (i.e., hadrons formed right after quark coalescence) to remove the effects from hadron scatterings and resonance decays. 
Figures~\ref{fig2}(d-f) show the results of this test at different randomization fraction $f$ after the normal parton freezeout at $\sigma_p=1.5$ mb in $pp$ events at $b=0$ fm. As expected, the parton $\langle \beta_{\perp} \rangle$ decreases linearly with the randomization fraction. 
More importantly, the same trend is also seen for the $\langle \mathrm{nsd} \rangle$ 
of both primordial $\rm pp +\bar{p}\bar{p}$ and $\rm BB +\bar{B}\bar{B}$ correlations, 
where a depression is seen in the normal AMPT-SM results (i.e., at $f=0$) but not seen when all freezeout partons have random azimuth in momentum. 

\begin{figure*}[htb]
\centering
\includegraphics[scale=0.8]{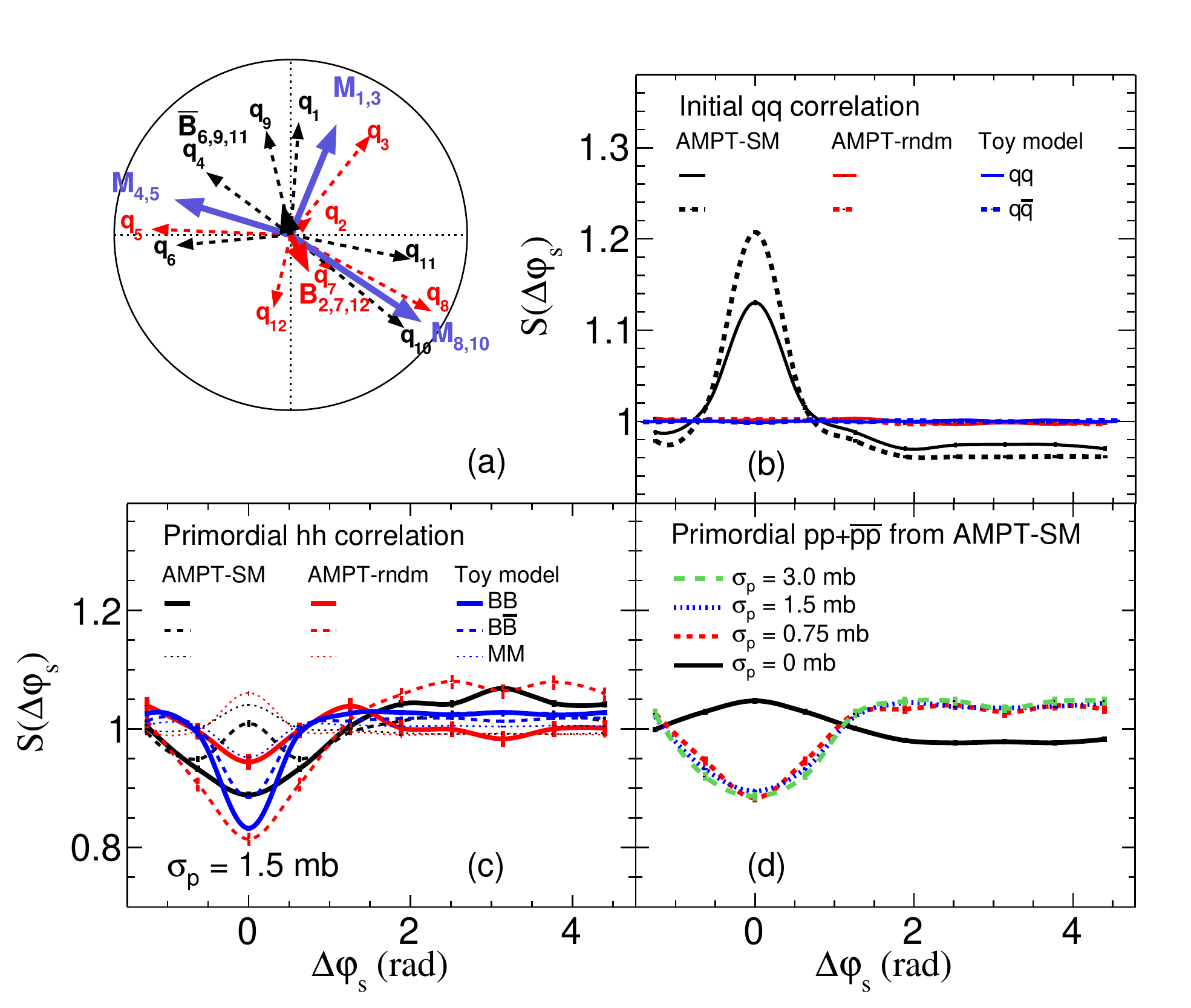}
\caption{(color online) (a) A toy model event in the transverse plane for the formation of a baryon, an antibaryon and three mesons from quark coalescence; (b) initial quark correlations in spatial azimuth $\varphi_s$ from normal AMPT-SM, the AMPT-SM model with random initial parton ($x,y$) positons (AMPT-rndm), and the toy model; (c) primordial two-baryon $\varphi_s$ correlations; (d) $\varphi_s$ correlations of primordial $\rm pp +\bar{p}\bar{p}$ from normal AMPT-SM at various parton cross sections.
}
\label{fig3}
\end{figure*}

We now construct a toy model to help understand how the two-hadron spatial correlations are affected by quark coalescence process. 
For simplicity, we consider events each containing a given equal number of quarks and antiquarks,  which are assumed to be randomly distributed inside a 2-dimensional (2D) unit circle in the transverse plane.
As an example, Fig.~\ref{fig3}(a) shows an event with 6 quarks (red dashed arrows) and 6 antiquarks (black dashed arrows), which form a baryon ($B_{2,7,12}$), an antibaryon ($\bar{B}_{6,9,11}$), and three mesons ($M_{1,3}$, $M_{4,5}$ and $M_{8,10}$). 
The quark coalescence is performed in the order of the parton number (1 to 12) and follows a 
procedure similar to this AMPT model~\cite{He:2017tla}. 
Specifically, for any available quark (antiquark), the toy model searches 
for the two closest quarks (antiquarks) in 2D relative distance as potential partners for forming a baryon (antibaryon) and also search for the closest antiquark (quark) 
as the potential partner for forming a meson. 
Then a baryon (antibaryon) or a meson is formed according to the coalescence criterion that depends on the coalescence distances and the $r_{BM}$ parameter~\cite{He:2017tla}. 
The solid arrow for each hadron in Fig.~\ref{fig3}(a) represents the average of the position vectors of its two or three coalescing partons.
From Fig.~\ref{fig3}(c), we see that the two-hadron correlations in the spatial azimuth $\varphi_s$ 
from the toy model (with 60 quarks and 60 antiquarks in each event) are not flat, although the quark-quark and quark-antiquark $\varphi_s$ correlations are flat by design as shown in Fig.~\ref{fig3}(b). However, the toy model also shows a near-side depression in baryon-antibaryon and meson-meson correlations, different from the normal AMPT-SM results.

We find that the initial quark-quark and quark-antiquark $\varphi_s$ correlations
from the normal AMPT-SM model are not flat at all, as shown in Fig.~\ref{fig3}(b).  
Therefore, we have also randomized the initial transverse positions ($x,y$) of partons before the parton cascade and obtained the AMPT-rndm results, 
which have flat initial two-parton $\varphi_s$ correlations like the toy model. 
We see in Fig.~\ref{fig3}(c) that the initial two-parton correlations indeed significantly affect 
the two-hadron correlations after quark coalescence, e.g., the baryon-antibaryon near-side  correlation changes from a weak peak in the AMPT-SM model to a strong depression in the AMPT-rndm model. There are still large differences between the two-hadron correlations from AMPT-rndm and the toy model although they have the same initial two-parton correlations; this is presumably because, unlike the AMPT model, the toy model does not consider parton momentum in its quark coalescence. Note that the parameter $r_{\rm BM}$ that controls the relative probability of coalescing a (anti)quark to a (anti)baryon or meson is set to $0.9$ in the toy model to reasonably describe the ratio of baryons to mesons, 
and adjusting the $r_{\rm BM}$ value will change the baryon/meson production ratio and the $\langle \mathrm{nsd} \rangle$ but not the general depression feature. 
In addition, the results shown in Fig.~\ref{fig3}(c) are for the parton cross section of 1.5 mb. 
The spatial azimuthal correlations of $\rm pp +\bar{p}\bar{p}$ from the normal AMPT model 
at various parton cross sections are shown in Fig.~\ref{fig3}(d), 
where the near-side correlation is a small bump before parton scatterings but 
becomes a depression after, similar to the momentum azimuthal correlations shown in Fig.~\ref{fig1}.

A physics picture for the near-side depression feature emerges out of these results: a parton matter is created and then expands to a finite volume, then the hadronization process (quark coalescence in the AMPT model for this study) converts parton degrees of freedom to primordial hadrons locally and produces multiple hadrons relatively close in the coordinate space including the 
spatial azimuth. 
The finite expansion of the parton matter creates a finite space-momentum correlation (often called the radial flow), and the transverse flow velocity of a local parton volume tends to change its hadrons from being close in spatial azimuth to being close in momentum azimuth.
Results from the AMPT-SM model with a finite parton cross section 
are qualitatively consistent with the experimental data. 
In addition, the PYTHIA model with string fragmentation alone, the AMPT-SM model with no parton expansion, and the AMPT-def model that is dominated by the hadron cascade 
all fail to produce the depression feature. 
Therefore, our study implies a finite parton expansion before hadronization, where a prerequisite is the existence of parton degrees of freedom, in $pp$ collisions $\rm \sqrt{s} =$ 7 TeV.

\section{Discussions}

\begin{figure*}[htb]
\centering
\includegraphics[scale=0.4]{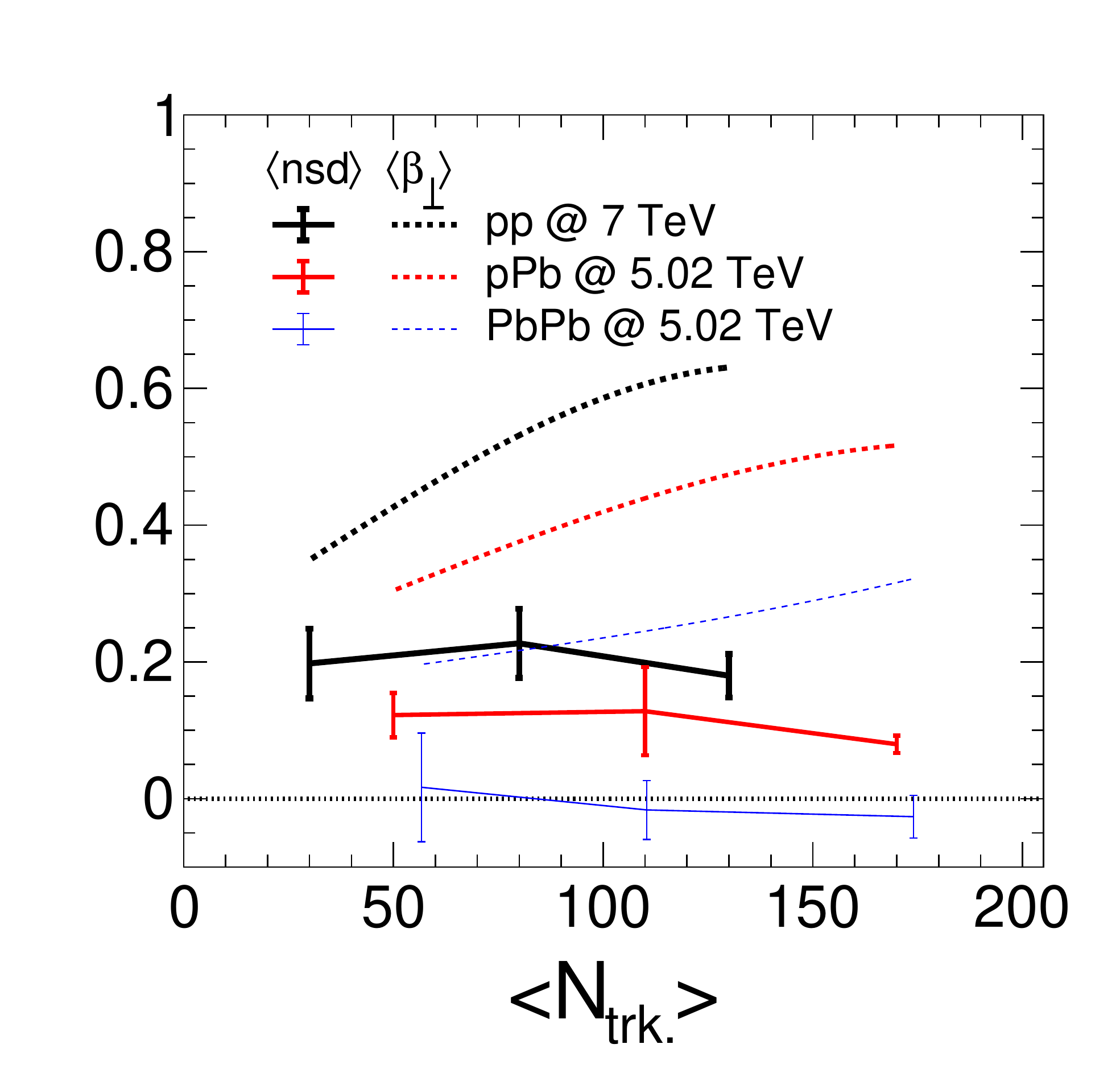}
\caption{(color online) The average near-side depression in $\rm pp + \bar{p}\bar{p}$ azimuthal correlations and the average transverse space-momentum correlation of freezeout partons at midrapidity versus the number of tracks in $pp$ collisions at $\mathrm{\sqrt{s}= 7}$ TeV at $\sigma_p$ = 1.5 mb, $p$Pb collisions and low-multiplicity Pb-Pb collisions at $\mathrm{\sqrt{s_{NN}}= 5.02}$ TeV at $\sigma_p$ = 3mb from the AMPT-SM model.
}
\label{fig4}
\end{figure*}

Regarding the depression feature in larger systems, Fig.~\ref{fig4} shows the average near-side depression $\langle \mathrm{nsd} \rangle$ of $\rm pp +\bar{p}\bar{p}$ correlation in $pp$ collisions at $\sqrt{s}$ = 7 TeV, $p$Pb, Pb-Pb collisions at $\rm \sqrt{s_{NN}}$ = 5.02 TeV as a function of $\rm N_{trk.}$ of charged particles within $|\eta|<3.0$ and $p_{T}<$ 5.0 GeV/c.  
The $\langle \mathrm{nsd} \rangle$ shows a weak dependence on $\rm N_{trk.}$ for a given collision system; on the other hand, its value at a given $\rm N_{trk.}$ clearly decreases with the collision system size from $pp$ to $p$Pb and then to low-multiplicity Pb-Pb collisions. 
We also see that the order of $\langle \mathrm{nsd} \rangle$ with the collision systems 
is the same as the order of the space-momentum correlation $\langle \mathrm{\beta_{\perp}} \rangle$. 
Because of the small initial volume, at a given $\rm N_{trk.}$ $pp$ collisions have the highest average number of collisions per parton (among the three collision systems) and the largest $\langle \mathrm{\beta_{\perp}} \rangle$ in the AMPT-SM model. 
Since a stronger space-momentum correlation converts the spatial azimuthal correlations more efficiently into the momentum azimuthal correlations, $pp$ collisions have the highest $\langle \mathrm{nsd} \rangle$ in Fig.~\ref{fig4}.
On the other hand, as $\rm N_{trk.}$ increases for a given collision system, 
the production of a larger number of baryons in each event tends to distribute the baryons more uniformly and thus dilute the depression~\cite{Zhang:2019bkf} despite the higher $\langle \mathrm{\beta_{\perp}} \rangle$. 
Therefore, the study of the two-hadron azimuthal near-side anticorrelation at LHC energies  
is the most promising for $pp$ collisions, which will be useful for the investigation  
of collectivity in small collision systems~\cite{Acharya:2020}. 

Calculations from the AMPT model using PYTHIA8 as the initial condition have explicitly shown~\cite{Zheng:2021} that the energy density at early times of $pp$ collisions at the LHC is much higher than the typical energy density at the QCD crossover phase transition. 
Therefore, we may expect theoretically that a partonic matter is created in such small systems at very high energies. 
However, our current results of $pp$ collisions from the AMPT model are consistent with the experimental data qualitatively but not quantitatively. As a result, they suggest the aforementioned physics picture but cannot be used to extract the value of the parton cross section, which is directly related to the properties such as the shear viscosity~\cite{Zhao:2020yvf}. 
Furthermore, the parton cross section directly determines the degree of equilibration of the partonic matter, 
where the typical value of 1.5-3 mb leads to a small average number of collisions per parton on the order of one 
~\cite{He:2015hfa,Li:2018leh}. 
In this case, the partonic matter may be far away from local equilibrium~\cite{He:2015hfa,Lin:2015ucn} and 
quite different from a quark-gluon plasma near equilibrium. 
Future improvements of the initial condition~\cite{Zheng:2021} and the hadronization criterion~\cite{Lin:2021mdn} would make the AMPT model more reliable for quantitative descriptions of the depression feature in $pp$ collisions. 
In addition, it will be useful to study the full azimuthal distribution, including the near-side correlation as well as multi-particle cumulants~\cite{Zhao:2021bef}, 
to better constrain the model and investigate the properties of the partonic matter.

\section{Conclusion}

To explore the physics origin of the near-side depression feature of baryon-baryon 
or antibaryon-antibaryon azimuthal correlations around $\Delta \varphi \approx 0$ in $pp$ collisions at $\rm \sqrt{s}=$ 7 TeV, a study with a multi-phase transport model 
has been performed using different parton cross sections and a free-streaming test.
The near-side correlation is not a depression without the expansion of the parton system. 
However, it becomes a depression with a finite parton expansion that leads to a finite space-momentum correlation of local partons at hadronization. 
A strong space-momentum correlation converts the spatial azimuthal correlations 
efficiently into the momentum azimuthal correlations. 
We have further investigated the effect of hadronization with a toy model for quark coalescence.  We find that the resultant two-hadron spatial azimuthal correlations in small systems have near-side structures such as a depression even though the initial two-parton spatial azimuthal correlations are flat; the initial parton-level spatial correlations are also found to be important. 
Therefore, we conclude that the initial parton-level correlations, quark coalescence and a finite expansion of the parton system are all important ingredients leading to the near-side depression. 
Our study suggests that a partonic matter with a finite expansion and lifetime is created in $pp$ collisions at $\rm \sqrt{s}=$ 7 TeV, especially given that the PYTHIA8 model with string fragmentation alone and AMPT model calculations with few or no parton rescatterings all fail to produce the depression feature. 
Whether the partonic matter is a quark-gluon plasma near local equilibrium or far away from local equilibrium requires further studies.

\section{Acknowledgements}

The work of L. Y. Zhang and J. H. Chen is supported in part by the Guangdong Major Project of Basic and Applied Basic Research No. 2020B0301030008,
the Strategic Priority Research Program of Chinese Academy of Sciences with Grant No. XDB34030000, and the National Natural Science Foundation of China under Contract Nos. 12025501, 11890710, 11890714 and 12147111. 
Z.-W. Lin is supported in part by the National Science Foundation under Grant No. PHY-2012947.

\end{document}